\documentclass[twocolumn,showpacs,preprintnumbers,amsmath,amssymb]{revtex4}

\usepackage{graphicx}
\usepackage{dcolumn}
\usepackage{bm}

\begin{document}


\title{Wave-particle duality and `bipartite' wave functions
for a single particle}
\author{Zeqian Chen}
\email{chenzq@mail.cnu.edu.cn}
\affiliation{%
Research Center for Mathematical Physics, School of Mathematical
Sciences, Capital Normal University, 105 North Road, Xi-San-Huan,
Beijing, China}
\affiliation{%
Wuhan Institute of Physics and Mathematics, Chinese Academy of
Sciences, 30 West District, Xiao-Hong-Shan, P.O.Box 71010, Wuhan,
China}

\date{\today}

\begin{abstract}
It is shown that `bipartite' wave functions can present a
mathematical formalism of quantum theory for a single particle, in
which the associated Schr\"{o}dinger's wave functions correspond to
those `bipartite' wave functions of product forms. This extension of
Schr\"{o}dinger's form establishes a mathematical expression of
wave-particle duality and that von Neumann's entropy is a
quantitative measure of complementarity between wave-like and
particle-like behaviors. In particular, this formalism suggests that
collapses of Schr\"{o}dinger's wave functions can be regarded as the
simultaneous transition of the particle from many levels to one. Our
results shed considerable light on the basis of quantum mechanics,
including quantum measurement.
\end{abstract}

\maketitle

Wave-particle duality, as manifest in the two-slit experiment,
provides perhaps the most vivid illustration of Bohr's
complementarity principle which refers to the ability of
quantum-mechanical entities to behave as waves or particles under
different experiment conditions \cite{Bohr}. Wave-like
(interference) behaviour can be explained by the superposition
principle \cite{Dirac}, while the usual explanation for the loss of
interference (particle-like behaviour) in a which-way experiment is
based on Heisenberg's uncertainty principle \cite{Feynman}. However,
it is demonstrated in the which-way experiment with an atom
interferometer that Heisenberg's position-momentum uncertainty
relation cannot explain the loss of interference for which the
correlations between the which-way detector and the atomic beams are
considered to be responsible \cite{DNR}. This explanation for
particle-like behaviour refers to quantum measurement and so is
inconsistent with the standard interpretation of quantum mechanics
\cite{M}. From a theoretical point of view, a measurement-free
express of wave-particle duality deserves research.

Recently, an differential equation for wave functions is proposed
for a single particle, which is equivalent to Schr\"{o}dinger's wave
equation and can be used to determine energy-level gaps of the
system \cite{Chen}. Contrary to Schr\"{o}dinger's wave equation,
this equation is on `bipartite' wave functions. It is shown that
these `bipartite' wave functions satisfy all the basic properties of
Schr\"{o}dinger's wave functions which correspond to those
`bipartite' wave functions of product forms. Here, we will show that
this extension of Schr\"{o}dinger's form establishes a mathematical
expression of wave-particle duality and that von Neumann's entropy
is a quantitative measure of complementarity between wave-like and
particle-like behaviors. It is concluded that a single particle
behaves like waves when it interfere with itself, while like
particle when entangle with itself. Thus, wave-particle duality is
just the complementarity of interference and entanglement for a
single particle. Further, from this formalism it is concluded that
collapses of Schr\"{o}dinger's wave functions can be regarded as the
simultaneous transition of the particle from many levels to one.
This sheds considerable light on quantum measurement.

Consider the quantum system of a single particle. Note that the
Hamiltonian for a single particle in an external field
is\begin{equation}\hat{H}(\vec{x}) = - \frac{\hbar^2}{ 2 m}
\nabla^2_{\vec{x}} + U(\vec{x} ),\end{equation}where
$\nabla^2_{\vec{x}} = \partial^2/\partial x^2_1 +
\partial^2/\partial x^2_2 + \partial^2/\partial x^2_3,$
$U(\vec{x})$ is the potential energy of the particle in the external
field, and $\vec{x} = (x_1, x_2, x_3) \in \mathbb{R}^3.$ The
Schr\"{o}dinger's wave equation describing dynamics of the particle
is\begin{equation}i\hbar \frac{\partial \psi (\vec{x}, t) }{\partial
t} = \hat{H}(\vec{x}) \psi (\vec{x}, t) = - \frac{\hbar^2}{ 2 m}
\nabla^2_{\vec{x}} \psi (\vec{x}, t) + U(\vec{x} ) \psi (\vec{x},
t).\end{equation}It is well known that the basis of the mathematical
formalism of quantum mechanics lies in the proposition that the
state of the particle can be described by a definite wave function
$\psi$ of Equation (2), whose stationary states determine its energy
levels. Moreover, the expectation value of an observable $\hat{Q}$
in the state corresponding to $\psi$ is determined by $\langle
\hat{Q} \rangle_{\psi} = \langle \psi | \hat{Q} | \psi\rangle.$

On the other hand, it is shown by the author \cite{Chen} that the
states of a single particle can also be described by definite
`bipartite' wave functions that obey the following
equation\begin{equation}i\hbar \frac{\partial \Psi (\vec{x},
\vec{y}; t) }{\partial t} = \left (\hat{H}(\vec{x}) -
\hat{H}(\vec{y}) \right ) \Psi (\vec{x}, \vec{y};
t),\end{equation}which is mathematically equivalent to
Schr\"{o}dinger's wave equation (2) and whose stationary states
determine the energy-level gaps of the system. In particular, the
expectation value of an observable $\hat{Q}$ in the state
corresponding to a definite `bipartite' wave function $\Psi$ is
determined by\begin{equation} \langle \hat{O} \rangle_{\Psi} =
\mathrm{Tr} \left [ \varrho^{\dagger}_{\Psi} \hat{O} \varrho_{\Psi}
\right ],\end{equation}where $\varrho_{\Psi}$ is an operator on
$L^2$ associated with $\Psi$ defined by $(\varrho_{\Psi} \varphi )
(\vec{x}) = \int \Psi (\vec{x}, \vec{y}) \varphi (\vec{y}) d^3
\vec{y}$ for every $\varphi \in L^2,$ provided $\Psi$ is normalized.
It is easy to check that if $\Psi (\vec{x}, \vec{y}) = \psi
(\vec{x}) \psi^* (\vec{y}),$ then $\langle \hat{O} \rangle_{\Psi} =
\langle \psi | \hat{O} | \psi \rangle.$ This concludes that our
expression Eq.(4) agrees with the interpretation of
Schr\"{o}dinger's wave functions for calculating expectation values
of any chosen observable.

The `bipartite' wave function $\Psi$ of physical meaning should be
`Hermitian':\begin{equation} \Psi^*(\vec{x}, \vec{y} ) =
\Psi(\vec{y}, \vec{x} ).\end{equation}Schr\"{o}dinger's wave
functions correspond to those `bipartite' wave functions of product
forms $\Psi (\vec{x}, \vec{y}) = \psi (\vec{x}) \psi^* (\vec{y}).$
Other `bipartite' wave functions that cannot be written as product
forms are `entangled' \cite{EPR}. By Schmidt's decomposition theorem
\cite{Schmidt}, for every normalized `bipartite' wave function $\Psi
(\vec{x}, \vec{y})$ there exist two orthogonal sets $\{\psi_n \}$
and $\{\varphi_n \}$ in $L^2_{\vec{x}}$ and $L^2_{\vec{y}}$
respectively, and a sequence of positive numbers $\{ \mu_n \}$
satisfying $\sum_n \mu^2_n = 1$ so that\begin{equation}\Psi
(\vec{x}, \vec{y}) = \sum_n \mu_n \psi_n (\vec{x}) \varphi^*_n
(\vec{y}).\end{equation}Then, we define the `entanglement' measure
of `bipartite' wave functions $\Psi$ by\begin{equation} S (\Psi) = -
\sum_n \mu^2_n \ln \mu^2_n.\end{equation}It is easy to show
that\begin{equation} S (\Psi) = - \mathrm{tr} \left [
\varrho_{\vec{x}} (\Psi) \ln \varrho_{\vec{x}} (\Psi) \right ] = -
\mathrm{tr} \left [ \varrho_{\vec{y}} (\Psi) \ln \varrho_{\vec{y}}
(\Psi) \right ] ,\end{equation}where $\varrho_{\vec{x}} (\Psi) =
\mathrm{tr}_{\vec{y}} \left ( | \Psi \rangle \langle \Psi | \right
)$ and $\varrho_{\vec{y}} (\Psi) = \mathrm{tr}_{\vec{x}} \left ( |
\Psi \rangle \langle \Psi | \right ).$ Hence, $S (\Psi)$ is the von
Neumann's entropy of the reduced density matrix $\varrho_{\vec{x}}
(\Psi)$ (or equivalently, $\varrho_{\vec{y}} (\Psi)$) of $| \Psi
\rangle \langle \Psi |$ (e.g., \cite{BDSW}). In the sequel, we will
show that $S (\Psi)$ is a quantitative measure of complementarity
between wave-like and particle-like behaviors.

Let us imagine a screen impermeable to electrons, in which two
slits, 1 and 2, are cut. We denote by $\psi_1$ the wave function of
an electron through slit 1 with slit 2 being covered, and $\psi_2$
the wave function of an electron through slit 2 with slit 1 being
covered. Then, the state of a single electron through slits 1 and 2
can be described by a `bipartite' wave function of
form\begin{equation}\begin{array}{lcl} \Psi (\vec{x}, \vec{y}) &=&
a_{11} \psi_1 ( \vec{x} ) \psi^*_1 (\vec{y}) + a_{12} \psi_1 (
\vec{x} ) \psi^*_2 (\vec{y})\\[0.4cm]&~&~~ + a_{21} \psi_2 ( \vec{x} ) \psi^*_1
(\vec{y}) + a_{22} \psi_2 ( \vec{x} ) \psi^*_2
(\vec{y}),\end{array}\end{equation}where $a^*_{ij} = a_{ji}$ and
$|a_{11}|^2 + |a_{12}|^2 + |a_{21}|^2 + |a_{22}|^2 =1.$ As following
are two special cases of Eq.(9):\begin{equation}\Psi_W (\vec{x},
\vec{y}) = \frac{1}{2} \left [  \psi_1 ( \vec{x} ) + \psi_2 (
\vec{x} ) \right ] \left [  \psi^*_1 ( \vec{y} ) + \psi^*_2 (
\vec{y} ) \right ],\end{equation} \begin{equation}\Psi_P (\vec{x},
\vec{y}) = \frac{1}{\sqrt{2}} \psi_1 ( \vec{x} ) \psi^*_1 ( \vec{y}
) + \frac{1}{\sqrt{2}} \psi_2 ( \vec{x} ) \psi^*_2 ( \vec{y}
).\end{equation}Accordingly, $\Psi_W$ corresponds to
Schr\"{o}dinger's wave function $\psi = \frac{1}{\sqrt{2}} (\psi_1 +
\psi_2),$ while there is no Schr\"{o}dinger's wave function
associated with $\Psi_P.$ A single electron described by $\Psi_W$
behaves like waves, while by $\Psi_P$ like particles. This is so
because for position, by (4) we have $\langle \hat{x}
\rangle_{\Psi_W} = | \psi_1 ( \vec{x}) + \psi_2 ( \vec{x})|^2 / 2$
and $\langle \hat{x} \rangle_{\Psi_P} = ( | \psi_1 ( \vec{x})|^2 + |
\psi_2 ( \vec{x})|^2 )/ 2,$ respectively. Generally, for every
$\Psi$ of Eq.(9) one has$$0 \leq S ( \Psi ) \leq S ( \Psi_P ) =
\frac{1}{2} \ln 2.$$When a single electron is described by $\Psi,$
the larger is $S ( \Psi ),$ more like particles it behaves
\cite{BSHMU}. Hence, $S ( \Psi )$ characterizes quantitatively
wave-particle duality for a single particle.

Let $\psi_n$ be the eigenfuncions of the Hamiltonian operator
$\hat{H},$ i.e., which satisfy the equation\begin{equation}
\hat{H}(\vec{x}) \psi_n (\vec{x}) = E_n \psi_n
(\vec{x}),\end{equation}where $E_n$ are the eigenvalues of
$\hat{H}.$ Since $\{ \psi_n (\vec{x}) \}$ is a complete orthogonal
set in $L^2_{\vec{x}},$ it is concluded that $\{ \psi_n (\vec{x})
\psi^*_m (\vec{y}) \}$ is a complete orthogonal set in
$L^2_{\vec{x}, \vec{y}}.$ Then, for every normalized `bipartite'
wave function $\Psi (\vec{x}, \vec{y})$ there exists a unique set of
numbers $\{ c_{n,m}\}$ satisfying $c^*_{n,m} = c_{m,n}$ and
$\sum_{n,m} | c_{n,m} |^2 = 1$ so that\begin{equation}\Psi (\vec{x},
\vec{y}) = \sum_{n,m} c_{n,m} \psi_n (\vec{x}) \psi^*_m
(\vec{y}).\end{equation}Note that$$ \left ( \hat{H}(\vec{x})-
\hat{H}(\vec{y}) \right ) (\psi_n (\vec{x}) \psi^*_m (\vec{y}) ) =
(E_n - E_m) \psi_n (\vec{x}) \psi^*_m (\vec{y}).$$Then, $c_{n,m}$ is
the probabilistic amplitude of the transition of the particle from
level $\psi_n$ to $\psi_m.$ The probability of getting $E_m$ on
measurement in a state with `bipartite' wave function $\Psi$
is\begin{equation}p_m = \sum_n | c_{n,m}|^2. \end{equation}In this
case, the collapse of $\Psi$ to $\psi_m$ can be regarded as the
simultaneous transition of the particle from levels $\psi_1, \psi_2,
\ldots$ to $\psi_m.$ The value of the associated change of energy of
the system is\begin{equation}\triangle E_m = \sum_n | c_{n,m}|^2 (
E_n - E_m ).\end{equation}Thus, our results suggest that von
Neumann's collapse of Schr\"{o}dinger's wave functions is just the
simultaneous transition of the particle from many levels to one
\cite{vN}. We therefore conclude that there are two basic change for
the system of a single particle, one is unitary, while the other is
von Neumann's collapse in such a sense that the simultaneous
transition of the particle from many levels (perhaps, only one) to
one. The simultaneous transition should play a role in quantum
measurement.

In conclusion, we show that `bipartite' wave functions can present a
mathematical formalism of quantum theory for a single particle. This
extends Schr\"{o}dinger's form of wave functions. It is presented a
mathematical expression of wave-particle duality and that von
Neumann's entropy is a quantitative measure of complementarity
between wave-like and particle-like behaviors. Our formalism
suggests that von Neumann's collapses of Schr\"{o}dinger's wave
functions is just the simultaneous transition of the particle from
many levels to one. Our results shed considerable light on the basis
of quantum mechanics, including quantum measurement.

This work was supported by the National Natural Science Foundation
of China under Grant No.10571176, the National Basic Research
Programme of China under Grant No.2001CB309309, and also funds
from Chinese Academy of Sciences.



\begin{thebibliography}{**}
\bibitem{Bohr}N.Bohr, {\it Albert Einstein: Philosopher-Scientist} (ed. Schilpp, P.A.)
200-241 (Library of Living Philosophers, Evanston, 1949).
\bibitem{Dirac}P.A.M.Dirac, {\it The Principles of Quantum
Mechanics} (Fourth edition, Oxford University Press, Oxford, 1958).
\bibitem{Feynman}R.P.Feynman and A.R.Hibbs, {\it Quantum Mechanics and
Path Integrals} (McGraw-Hill, Inc., 1965).
\bibitem{DNR}S.D\"{u}rr, T.Nonn, and G.Rempe, Nature
{\bf 395}, 33(1998).
\bibitem{M}P.Mittelstraedt, {\it The Interpretation of Quantum
Mechanics and the Measurement Process} (Cambridge University Press,
Cambridge, 1998).
\bibitem{Chen}Z.Chen, quant-ph/0604083.
\bibitem{EPR}A.Einstein, B.Podolsky, and N.Rosen, Phys. Rev. {\bf
47}, 777(1935); E.Schr\"{o}dinger, Naturwissenschaften {\bf 23},
807(1935); R.Werner, Phys.Rev.A {\bf 40}, 4277(1989).
\bibitem{Schmidt}E.Schmidt, Math.Ann.{\bf 63}, 433(1907).
\bibitem{BDSW}C.H.Bennett, D.P.DiVincenzo, J.Smolin, and
W.K.Wootters, Phys.Rev.A {\bf 54}, 3814(1996).
\bibitem{BSHMU}E.Buks, S.Schuster, M.Heiblum, D.Mahalu, and V.Umansky, Nature
{\bf 391}, 871(1998).
\bibitem{vN}J.von Neumann, {\it Mathematical Foundations of Quantum Mechanics}
(Princeton University Press, Princeton, 1955).
\end{thebibliography}
\end{document}